\documentclass[amsmath,amssymb,pra,aps,showpacs,superscriptaddress,twocolumn]{revtex4-2}
\usepackage{amsmath,amsfonts,amssymb,amsthm,graphics,graphicx,epsfig,bbm}
\usepackage[colorlinks=true,citecolor=blue,linkcolor=blue,urlcolor=blue]{hyperref}
\usepackage[usenames]{color}
\usepackage{graphicx}
\usepackage{subfigure}
\usepackage{amsmath}
\usepackage{epsfig}
\usepackage{dcolumn}
\usepackage{bm}
\usepackage{color}
\usepackage{times}
\usepackage{epstopdf}
\usepackage{amssymb}
\usepackage{amstext}
\usepackage{latexsym}
\usepackage{hyperref}
\usepackage{amsfonts}
\usepackage{psfrag}
\usepackage{soul,xcolor}
\usepackage[normalem]{ulem}
\usepackage{dsfont}
\usepackage{txfonts}
\usepackage{float}
\usepackage{makecell}
\usepackage{comment}
\usepackage{appendix}
\usepackage{diagbox}

\usepackage{graphicx}
\usepackage{dcolumn}
\usepackage{bm}

\newcommand{\ket}[1]{\vert #1 \rangle}

\newcommand{\Tr}{\mathrm{Tr}}

\newcommand{\XC}{\textcolor{black}}

\begin{document}

\title{Exploring Ground States of Fermi-Hubbard Model on Honeycomb Lattices with Counterdiabaticity}

\author{Jialiang Tang}
\affiliation{Department of Physical Chemistry, University of the Basque Country UPV/EHU, Apartado 644, 48080 Bilbao, Spain}

\author{Ruoqian Xu}
\affiliation{Department of Physical Chemistry, University of the Basque Country UPV/EHU, Apartado 644, 48080 Bilbao, Spain}

\author{Yongcheng Ding}
\affiliation{Department of Physical Chemistry, University of the Basque Country UPV/EHU, Apartado 644, 48080 Bilbao, Spain}
\affiliation{Institute for Quantum Science and Technology, Department of Physics, Shanghai University, Shanghai 200444, China}

\author{Xusheng Xu}
\affiliation{Department of Physics, State Key Laboratory of Low-Dimensional Quantum Physics, Tsinghua University, Beijing 100084, China}

\author{Yue Ban}
\affiliation{Departamento de F\'isica, Universidad Carlos III de Madrid, Avda. de la Universidad 30, 28911 Legan\'es, Spain}

\author{Manhong Yung}
\affiliation{Department of Physics, Southern University of Science and Technology, Shenzhen 518055, People's Republic of China}
\affiliation{Shenzhen Institute for Quantum Science and Engineering, Southern University of Science and Technology, Shenzhen 518055, People's Republic of China}
\affiliation{Guangdong Provincial Key Laboratory of Quantum Science and Engineering, Southern University of Science and Technology, Shenzhen 518055, People's Republic of China}
\affiliation{Shenzhen Key Laboratory of Quantum Science and Engineering, Southern University of Science and Technology, Shenzhen 518055, People's Republic of China}

\author{Axel Pérez-Obiol}
\affiliation{Departament de Física, Universitat Autònoma de Barcelona, E-08193 Bellaterra, Spain}

\author{Gloria Platero}
\affiliation{Instituto de Ciencia de Materiales de Madrid (CSIC), Cantoblanco, E-28049 Madrid, Spain}

\author{Xi Chen}
\email{xi.chen@ehu.eus}
\affiliation{Department of Physical Chemistry, University of the Basque Country UPV/EHU, Apartado 644, 48080 Bilbao, Spain}
\affiliation{EHU Quantum Center, University of the Basque Country UPV/EHU, Barrio Sarriena, s/n, 48940 Leioa, Biscay, Spain}

\begin{abstract}
Exploring the ground state properties of many-body quantum systems conventionally involves adiabatic processes, alongside exact diagonalization, in the context of quantum annealing or adiabatic quantum computation. Shortcuts to adiabaticity by counter-diabatic driving serve to accelerate these processes by suppressing energy excitations. Motivated by this, we develop variational quantum algorithms incorporating the auxiliary counter-diabatic interactions, comparing them with digitized adiabatic algorithms. These algorithms are then implemented on gate-based quantum circuits to explore the ground states of the Fermi-Hubbard model on honeycomb lattices, utilizing systems with up to 26 qubits. The comparison reveals that the counter-diabatic inspired ansatz is superior to traditional Hamiltonian variational ansatz. Furthermore, the number and duration of Trotter steps are analyzed to understand and mitigate errors. Given the model’s relevance to materials in condensed matter, our study paves the way for using variational quantum algorithms with counterdiabaticity to explore quantum materials in the noisy intermediate-scale quantum era.
\end{abstract}

\maketitle


\section{\label{sec:level1}INTRODUCTION}

Efficiently simulating complex quantum systems stands as a pivotal capability of quantum computers ~\cite{2019_Smith_npj, Iulia2009Science, Chris2020PRB, Dave2015PRA, Stanisic2022NC}, often regarded as a potent tool for investigating quantum materials~\cite{Nathalie_2021_Nature, Ma_2020_npj, Bela_2020_Chemical, ObiolPRA2022}, chemistry~\cite{Lanyon_2010_Nature, Yudong_2019_Chem, Sam_2020_RMP}, biology~\cite{Fedorov2021_NC, Marx2021_NC}, and nuclear physics~\cite{nuclear}, all of which are governed by the principles of quantum mechanics. This proposal has transitioned from a conceptual idea to a practical endeavor with the development of experimental technologies in the noisy intermediate-scale quantum (NISQ) era. Among the most celebrated algorithms in this context is the variational quantum algorithm (VQA)~\cite{Cerezo2021NRP, Wecker2015PRA, Lennart2021PRL, Tyson2019PRA}, which aims to reduce circuit depth by integrating quantum computers with classical optimizers. The circuit utilizes parameterized blocks of quantum gates as ansatz, tailored for different scenarios. Hardware-efficient ansatz address hardware limitations ~\cite{Kandala2017Nature}, while unitary coupled clusters are optimized for quantum chemistry and condensed matter systems ~\cite{Romero2018QST, Shen2017}. These ansatz, derived from the system's Hamiltonian, form the core of Variational Quantum Algorithms (VQAs), facilitating efficient exploration and optimization of quantum states. A straightforward application of VQA is in ground state searching, a critical aspect for quantum information processing and many-body physics. Exact diagonalization, while theoretically accurate, suffers from exponentially growing computational complexity, making it impractical for large-scale systems. However, this complexity can be mitigated by employing variational circuits to approximate the ground state through energy expectation minimization. This underscores the requirement for designing an adequate ansatz to facilitate finding the ground state, while increasing expressibility thus potentially reducing the impact from the notorious barren plateau caused by hardware-efficient ansatz, large system size, and deep circuits~\cite{McClean2018NC, Pesah2021PRX}.

In accordance with the adiabatic theorem, a system is expected to evolve while preserving its ground state integrity as long as the adiabatic criteria remain unviolated~\cite{RevModPhys}. This ansatz mirrors the structure observed in digital adiabatic quantum computing, reminiscent of the quantum-approximated optimization algorithm (QAOA), which seeks to identify optimal annealing schedules~\cite{Chandarana2022PRR}. 
This naturally leads to the assumption that one can introduce counterdiabaticity in the design of ansatz to enhance the performance of VQAs, akin to the approach seen in adaptive derivative-assembled problem tailored (ADAPT) VQAs \cite{Sophia1, Sophia2}. This incorporation of counterdiabaticity aims to cancel induced energy excitation \cite{Rice, Berry}, effectively serving as shortcuts to adiabaticity~\cite{Chen2010PRL, Odelin2019RMP}. Most of these terms, involving many-body interactions, are difficult or even impossible to directly realize in realistic quantum systems, due to their complexity. However, this limitation no longer exists in ansatz design due to gate decomposition. By incorporating counter-diabatic (CD) interactions into VQAs, one can significantly improve performance at the same energetic cost or gate numbers, as recently demonstrated through rigorous benchmarking~\cite{Xu2024Arxiv}. These digitized counter-diabatic quantum algorithms (DCQAs) have successfully determined the ground state of simpler many-body models and have found application in interdisciplinary fields such as molecular docking~\cite{Ding2023Arxiv} and protein folding~\cite{Pranav2023PRA}, showcasing advantages over conventional AQAs~\cite{Narendra2021PRApplied}.

In this article, we propose DCQAs to explore the ground state of the Fermi-Hubbard (FH) model on a honeycomb lattice. The extension beyond other lattice geometries introduces additional geometric complexity and distinctive electronic properties. Following the model description, we utilize DCQAs to solve for its ground state energy, and compare their performance against digital adiabatic algorithms aided by counterdiabaticity. Our results include a comprehensive analysis of Trotter error, adiabatic error, and ground state properties. Additionally, for larger system sizes, we simulate VQAs with up to 26 qubits, assessing the performance of both Hamiltonian variational ansatz and CD-inspired ansatz. Our results provide the possible feasibility of utilizing quantum computing in many-body physics and quantum materials research.

\section{\label{sec:level-model} Preliminaries on model and hamiltonian}

\begin{figure}[ht]
\centering
\includegraphics[scale=0.42]{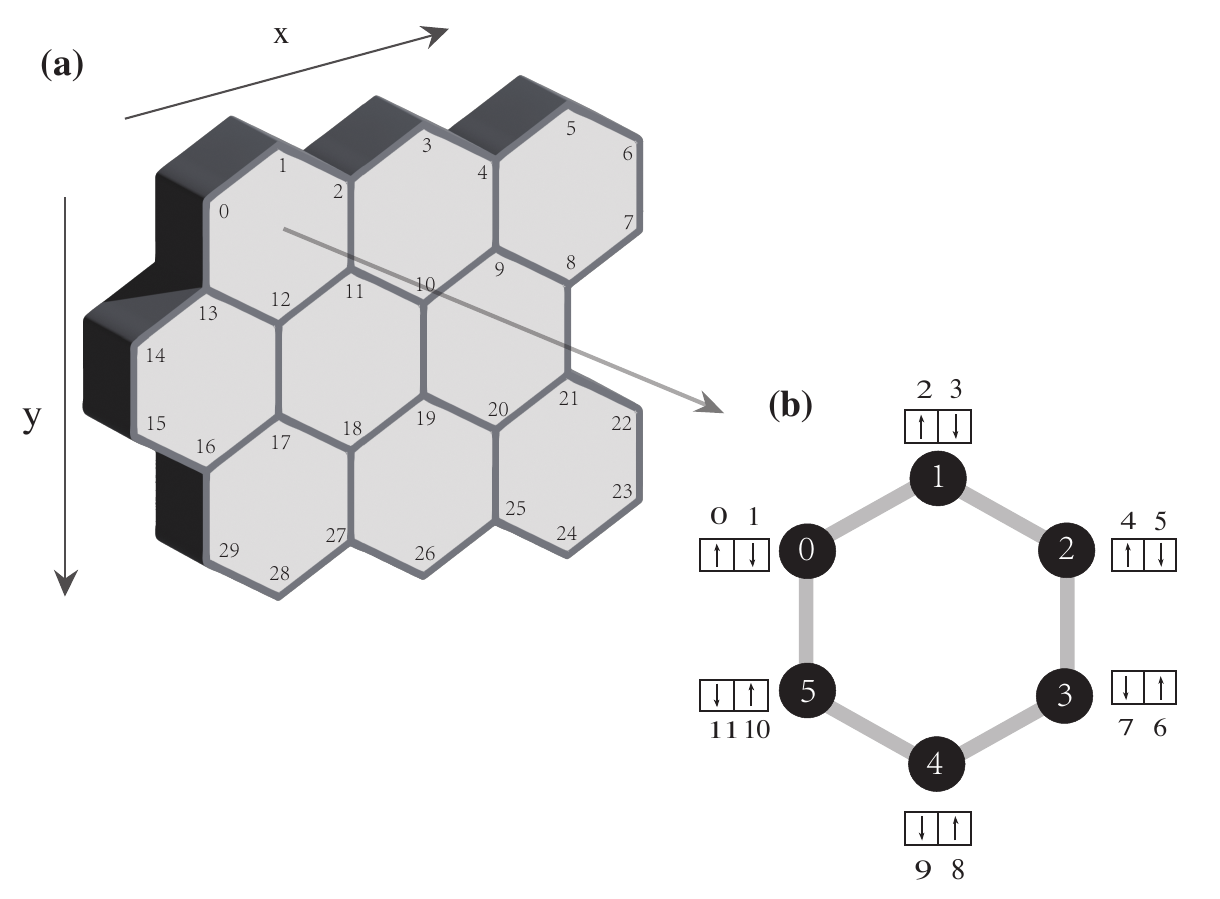}
\caption{Schematic diagram of the FH model: regular hexagonal honeycomb structure is used to describe the lattices. (a) The $x$ axis represents the horizontal direction and the $y$ axis represents the vertical direction. Each node denotes a lattice site and they are ordered in zig-zag. (b) For each lattice, black dots represent site numbers with spin up and down, which needs one qubit. Spins are labeled with qubit numbers as denoted here.}
\label{fig1}
\end{figure}

The FH model is a celebrated model in condensed matter physics for characterizing how electron interactions influence the behavior of quantum materials~\cite{Esslinger2010FermiHubbardPW}. When extended to honeycomb lattices, this model introduces distinctive geometric aspects, offering the theoretical framework to study electronic phenomena and quantum phases in materials exhibiting hexagonal symmetry.  By employing quantum algorithms, we can simulate this model on quantum computers, probing its ground state that encodes crucial information regarding the stable electronic arrangement and intrinsic properties of quantum materials within this geometric context. 
The system's Hamiltonian reads
\begin{equation}
\label{eq:h_fh}
\hat{{H}}_{\text{FH}} =  -\tau \sum_{\langle i,j\rangle, \sigma}c_{i,\sigma}^{\dagger} c^{\textcolor{white}{\dagger}} _{j,\sigma} + U\sum_i c_{i,\uparrow}^\dagger
c_{i,\uparrow}^{\textcolor{white}{\dagger}} c_{i,\downarrow}^\dagger c_{i,\downarrow}^{\textcolor{white}{\dagger}},
\end{equation}
where $ \hat{{H}}_{\text{h}} = -\tau \sum_{\langle i,j\rangle, \sigma}c_{i,\sigma}^{\dagger} c^{\textcolor{white}{\dagger}} _{j,\sigma}$ 
and  $ 
\hat{{H}}_{\text{c}}= U\sum_{i} c_{i,\uparrow}^\dagger
c_{i,\uparrow}^{\textcolor{white}{\dagger}} c_{i,\downarrow}^\dagger c_{i,\downarrow}^{\textcolor{white}{\dagger}}
$ are hopping and Coulomb terms respectively. Here, the notation $\langle i,j \rangle$ restricts adjoint sites on lattice, spin $\sigma\in\{\uparrow,\downarrow\}$, $c_{i,\sigma}^\dagger~(c_{i,\sigma}^{\textcolor{white}{\dagger}})$ denotes creation (destruction) of an electron on site $i$ with the spin $\sigma$, $\tau$ and $U$ are respectively the hopping and Coulomb coupling coefficient.
The sub-Hamiltonian is explained as follows: the hopping term $\hat{H}_{\text{h}}$ describes the action when an electron hops from site $i$ to site $j$ while keeping the spin $\sigma$. The Coulomb term $\hat{H}_{\text{c}}$ characterizes the Coulomb potential generated by electrons occupying the same site. The Hamiltonian for hopping $\hat{H}_{\text{h}}$ includes both horizontal and vertical terms, as depicted in Fig.~\ref{fig1}(a) along the $x$ and $y$ axes. The lattice is arranged in a honeycomb pattern, with each site possessing two orbitals: one for spin up and the other for spin down. The hexagons are all regular and arranged in a zig-zag configuration. 

Until now, various methods have been proposed to map fermionic operators onto quantum computers, among which the Jordan-Wigner (JW) and Bravyi-Kitaev (BK) mappings stand out, for transformation of electronic states and operators to states of and operations upon
qubits in hybrid classical-quantum algorithms \cite{Jordan, Bravyi, Seeley, Tranter}. For our proposal, the JW mapping is a favorable choice as it satisfies the canonical anticommutation relations of fermionic operators, $\{a_i^{\textcolor{white}{\dagger}}, a_j^\dagger\}=\delta_{ij}$, where $i,j$ are arbitrary sites in the honeycomb lattice, and it is cost-effective and feasible in the scalability of systems with dozens of qubits. The constructed qubit operators and the fermionic operators need to satisfy the following relation given by the JW transformation method: 
\begin{align}
c_{j}^\dagger &\mapsto \frac{1}{2}\left(X_{j} - iY_{j}\right)Z_{1}\cdots Z_{j-1}, \\
c_{j} &\mapsto \frac{1}{2}\left(X_{j} + iY_{j}\right)Z_{1}\cdots Z_{j-1}, 
\end{align}
where $c_{j}^\dagger~(c_{j} )$ is the creation (destruction) operator of fermions, $X_{j},~Y_{j},~Z_{j}$ are Pauli operators.

In the honeycomb lattice with a single-layer structure, the orientation can be zig-zag to match the qubit and site numbers required for a comprehensive lattice description. Additionally, two qubits are assigned to each lattice point, as shown in Fig.~\ref{fig1}(b), corresponding to the two orbitals present, namely spin up or down. For instance, fermions possessing spin up and down on site number $2$ correspond to qubit numbers $4$ and $5$, respectively.


\begin{figure*}[t]
\centering
\includegraphics[scale=0.48]{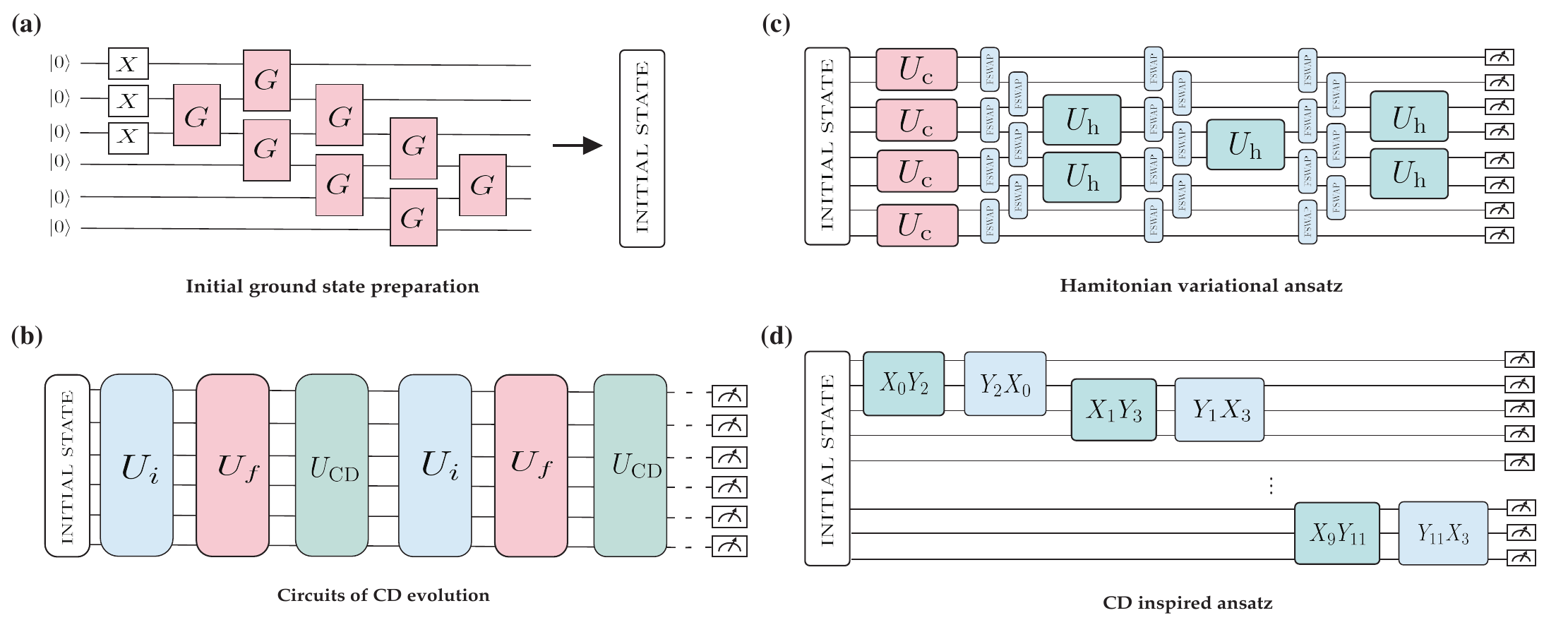}
\caption{Quantum circuits and ansatz structure. (a) An example of how to prepare initial ground state in a six-qubit quantum circuit, where $X$ gates are used to create fermions on the first three spin orbitals and Givens rotation gates are parameterized gates used to prepare the state to the corresponding initial state. (b) CD evolution circuits, where $U_i = \exp{(-iH_{\text{h}}t/\hbar)}$, $U_f = \exp{(-iH_{\text{FH}}t/\hbar)}$,  and $U_{\text{CD}} = \exp{(-iH_{\text{CD}}t/\hbar)}$ act alternately on the initial state and measure gates are put at the end of circuits so that errors can be estimated. (c) Hamitonian variational ansatz: $U_{\text{c}}$, $U_{\text{h}}$, and ${\rm FSWAP}$ are Coulomb, hopping, Fermi-SWAP operator. (d) CD-inspired ansatz, only two-body interactions are implemented in the circuits. Details of these local operators are summarized in Appendix~\ref{app: cd pool}.}
\label{fig2}
\end{figure*}

\section{\label{sec:level3} Speeding up Adiabatic algorithm}

There exist many algorithms successfully implemented on the strongly correlated electron system. In this section, we initially evaluate the performance of digitized AQAs, complemented by CD driving, serving as a reference for subsequent comparisons with DCQAs. Both approaches begin with the preparation of the initial state as the ground state of a suitable mixing Hamiltonian. Subsequently, we digitize the evolution of the Hamiltonian using quantum circuits. Finally, we measure the expectation values of joint spin operators to reconstruct the system's energy at the final time step.

\subsection{Initial state preparation}
\label{sec:state_pre}

The initial state for adiabatic evolution is chosen to be the ground state of $\hat{H}_{\text{h}}$ which is a fermionic Gaussian state and can be efficiently prepared on a quantum computer. As shown in Fig.~\ref{fig2}(a), we start from the occupied orbits with the lowest energy, half-full configuration. $X$ gates need to be applied on half of the spin-orbits to create one electron in each orbital. Then Givens rotation gates can be adjusted to prepare the initial state \cite{McClean2020QST}.
In the context of a honeycomb lattice, the complexity of preparing the initial state scales as $\mathcal{O}(N^2_{\text{site}})$, where $N_{\text{site}}$ represents the number of lattice sites. As the lattice size increases, the computational complexity becomes more closely associated with the number of qubits required, scaling as $\mathcal{O}(4N^2_{\text{q}})$, indicating a quadratic increase in complexity with respect to the number of qubits.

\begin{figure*}[t]
\centering
\includegraphics[scale=0.46]{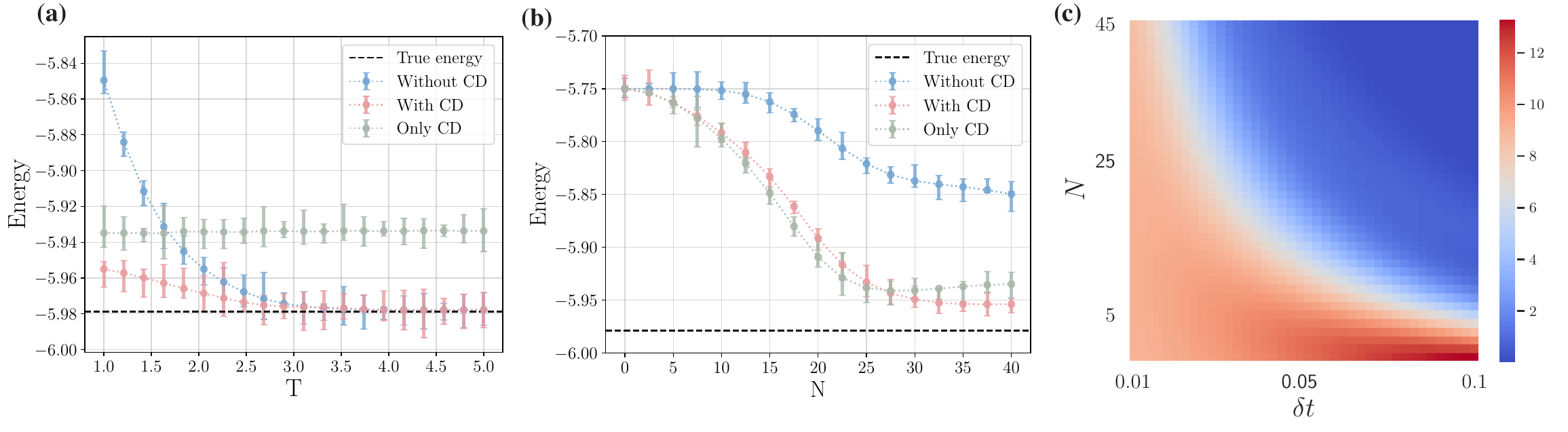}
\caption{Energy as a function of evolution time (T) and Trotter steps (N) for $1\times 1$ lattice. The blue and pink dash lines are the average of 10 measurements, while the black dash line is the ground energy. (a): Ten random initial parameters are chosen then the system evolves with CD, without CD and only CD with respect to time.  (b): Ten sets of random parameters are chosen to be the initial and energy changes with Trotter step $\text{N}$.  $\delta t=0.02$. (c): Energy error $\Delta E$ is a function of Trotter step (N) and Trotter time ($\delta t$). The colour bar displays the error difference between adiabatic evolution with and without CD interactions, defined in Eq.~\eqref{error-bar}. The other parameters: shots=$30000$, $\tau=1$ and $U=1.5$.}
\label{fig3}
\end{figure*}

\subsection{Adiabatic evolution and its acceleration}

For an adiabatic quantum process, the system gradually evolves from the ground state of the initial Hamiltonian to that of the final Hamiltonian. The Hamiltonian is given by
\begin{equation}
\label{eq:adiabatic h}
\hat{H}_a(t) = [1-\lambda(t)]\hat{H}_{i} + \lambda(t)\hat{H}_f, 
\end{equation}
where $\hat{H}_{i}$ and $\hat{H}_{f}$ represent the initial and final Hamiltonian, respectively. The scheduling function $\lambda(t)$ varies within the interval $[0,1]$ when $t\in$ [0, T], where T is the total evolution time. According to the adiabatic theorem, if T is sufficiently large (inversely proportional to the energy gap), the system will remain in its instantaneous eigenstate. The choice of $\hat{H}_{i}$ is crucial, and it is typically selected based on the ease of constructing its ground state on a quantum computer. In the case of the FH model, the initial state $\hat{H}_{i}$ is the ground state of the hopping term Hamiltonian $\hat{H}_{\text{h}}$~\cite{Jiang2018PRA} described in Eq.~\eqref{eq:h_fh}. On the other hand, $\hat{H}_{f}$ corresponds to the full FH Hamiltonian $\hat{H}_{\text{FH}}$, as explained in Eq.~\eqref{eq:h_fh}. By defining $\hat{H}_{i} = H_{\text{h}}$ and $\hat{H}_{f} = H_{FH}$, the Hamiltonian (\ref{eq:adiabatic h}) can be expressed as
\begin{equation}
\hat{H}_a(t) = \hat{H}_{\text{h}} + \lambda(t)\hat{H}_{\text{c}}.
\end{equation}

However, adiabatic quantum computing is inherently limited by its temporal demands, rendering the system vulnerable to noise and decoherence. To mitigate this,  CD driving \cite{Rice} (or equivalently transitionless algorithm \cite{Berry}), one of shortcut-to-adiabaticity techniques \cite{Odelin2019RMP}, can suppress non-adiabatic transitions, yielding the following Hamiltonian \cite{Narendra2021PRApplied}:
\begin{equation}
\hat{H}_{\text{tot}} = \hat{H}_a(t) + \hat{H}_{\text{CD}},
\label{eq:cd}
\end{equation}
where $\hat{H}_{\text{CD}} = \dot{\lambda}{\hat{A}}_\lambda$, with ${\hat{A}}_\lambda$ being the gauge potential. While there are various techniques to compute the CD Hamiltonian, obtaining exact CD terms for many-body systems is challenged by exact diagonalization. To address this, the nested commutator (NC) method~\cite{Sels, Pieter2019PRL} is proposed to approximate CD term. The gauge potential approximation reads
\begin{equation}
\hat{A}_\lambda^{(l)} = i\sum_{k=1}^l\alpha_k(t)\mathop{\underbrace{[\hat{H}_a, [\hat{H}_a,\cdots[\hat{H}_a}}\limits_{2k-1}, ~ \partial_\lambda \hat{H}]]],
\end{equation}
where $l$ denotes the order of the NC expansions, and becomes the exact CD term when $l \rightarrow \infty$. After minimizing the action $S = \Tr[G^2]$, where $G = \partial_\lambda \hat{H} - i[\hat{H}, \hat{A}_\lambda]$, we determine the coefficient $\alpha_{k}(t)$ for achieving the gauge potential.

To implement the evolution of quantum circuits, we utilize the first-order Trotter-Suzuki formula. The unitary evolution operator can be discretized into N steps with a timestep length of 
$\delta t$ (see Appendix~\ref{app: digital}). The quantum circuits for implementing adiabatic algorithms with CD terms involved are illustrated in Fig.~\ref{fig2}(b). 
In quantum computing, an optimal scenario involves a substantial value for Trotter steps N. However, increased quantum circuit depth also increases susceptibility to gate error and decoherence. In addition, the errors in the Trotter approximation become more pronounced  when 
N assumes a relatively small value.

\subsection{Measurement}

Ideally, obtaining detailed information about the quantum state, such as through quantum state tomography, allows for the calculation of the expected value of the Hamiltonian to determine the system's energy. In the context of the FH model, this involves reconstructing the expected value from the measured probabilities of the quantum state under newly defined basis vectors, achieved by diagonalizing each sub-Hamiltonian.

In the JW transformation, the Coulomb interaction terms are represented by matrices in the form $\frac{1}{4}(I-Z_i)(I-Z_{i+1})$, where the diagonal matrix elements can be measured directly in the computational basis. The expectation value of Coulomb terms is determined by the sum of probabilities for each site being in the state $\ket{11}$. The commutativity of all Coulomb terms allows for their simultaneous measurement.

For hopping terms, particularly vertical hopping terms, measuring the corresponding Pauli string of qubits, represented by $\frac{1}{2}(X_iX_j + Y_iY_j)Z_{i+1}\cdots Z_{j-1}$, poses a challenge due to its complexity. The Fermi-SWAP gate is utilized to transform from non-neighbouring qubits into nearest neighbouring ones, thereby eliminating the unnecessary Pauli $Z$ operators in the string. To measure the hopping energy between qubits $i$ and $j$, the Hamiltonian $\frac{1}{2}(X_iX_j + Y_iY_j)$ is diagonalized using a few quantum gates \cite{ObiolPRA2022}. The measurement outcome is then expressed as the probability difference between the states $\ket{01}$ and $\ket{10}$. 

In the study of a $1 \times 1$ honeycomb lattice, determining the total energy involves summing probabilities from four distinct groups of measurements. The first group captures Coulomb terms, representing the local energy contributions at individual lattice sites. The next two groups encompass horizontal hopping terms, with the first set covering qubit pairs $(0,2),~(1,3),~(6,8),~(7,9)$ and the second set involving pairs $(2,4),~(3,5),~(8,10),~(9,11)$. These pairs represent the kinetic energy contributions from electrons hopping between adjacent sites horizontally. The fourth group targets vertical hopping terms involving pairs $(4,7),~(5,6),~(0,11),~(1,10)$, which account for vertical interactions between sites. To accurately reconstruct the lattice's total energy, simulations have conducted using the adiabatic evolution with and without CD driving, in which measurements are taken for $30,000$ times at each T and N for $10$ instances. This extensive measurement strategy aims to mitigate statistical fluctuations and enhance the precision of the energy calculation, suggesting that even more measurements might be necessary for further accuracy.

\subsection{Simulation and error analysis}

Due to limitations in quantum resources, simulating large-scale grid models using adiabatic algorithms poses challenges.
The simulation results discussed here focus on estimating the ground energy of a $1\times 1$ honeycomb lattice, utilizing quantum circuits through adiabatic algorithms with and without CD interactions. The simulation begins from the ground state of an initial Hamiltonian, denoted by $|\psi_{i}\rangle$, which corresponds to the hopping term Hamiltonian $\hat{H}_{\text{h}}$, and aims to reach the ground state of the full FH Hamiltonian $\hat{H}_{\text{FH}}$, represented by $|\psi_f\rangle$.

The annealing schedule used here is $\lambda(t)=\sin^2[\frac{\pi}{2}\sin^2(\frac{\pi t}{2T})]$ \cite{Narendra2021PRApplied}, which modulates the interpolation between the initial Hamiltonian $\hat{H}_{\text{h}}$ and the final Hamiltonian $\hat{H}_{\text{FH}}$ over the evolution time T. CD terms, incorporating additional terms to suppress excitation that deviate from the ground state, demonstrate a significant advantage by achieving the ground energy in a shorter time. As shown in Fig.~\ref{fig3}(a), this advantage is particularly profound before the time $\rm{T<2.5}$, where the CD terms effectively prevent excitation energy from the ground one.  The analysis specifically considers first-order NC CD terms, including two-body and many-body interactions, obtained by the method detailed in Appendix \ref{app: cd pool}.
In Fig.~\ref{fig3}(b), with the time step $\delta t = 0.02$ and total evolution time $\rm{T=1}$, increasing the Trotter number results in a remarkable reduction of ground energy for counter-adiabatic evolution. This suggests that incorporating higher-order CD terms could further refine the ground energy estimation of the FH model, albeit at the cost of increased circuit complexity. The error bars noted in both Fig.~\ref{fig3}(a) and (b) reflect the statistical uncertainty arising from multiple measurements, emphasizing the probabilistic nature of quantum measurements in these simulations.

To further justify the advantage, we analyze the errors for digitized adiabatic algorithms in terms of various $\delta t$ and N. 
We define the error by the distance from ground energy to the estimated energy, expressed as:
\begin{equation}
\label{error-bar}
\Delta E = \left| \frac{\langle \psi_{g}|\hat{H}_{\text{FH}}|\psi_{g}\rangle-\langle\psi(\rm{T})|\hat{H}_{\text{FH}}|\psi(\rm{T})\rangle}{\langle \psi_{g}|\hat{H}_{\text{FH}}|\psi_{g}\rangle - \langle\psi_i|\hat{H}_{\text{FH}}|\psi_i\rangle}\right|
\times 100\%.
\end{equation}
where $|\psi_g\rangle$ and $|\psi_i\rangle$ are the ground state of the FH model and initial Hamiltonian, respectively, and $|\psi(\rm{T})\rangle$ represents the evolved state at $\rm{T}= \rm{N} \times \delta t$ in the circuit, depending on the Trotter step N and Trotter time $\delta$. 
Fig.~\ref{fig3}(c) illustrates the relative errors for digitized adiabatic evolution and its improvement by CD interactions. 
The advantage of CD driving is apparent when $\rm{T}$ is relatively small, and both approaches can approximate the ground state energy effectively when T is relatively large. However, it is notable that with fewer Trotter steps ($\rm{N}=5$), the adiabatic error is larger, especially when $\delta t=0.1$. In contrast, the CD-assisted evolution demonstrates better suppression of the Trotter error. Indeed, the results are consistent with the analysis of digitized CD driving, demonstrating its advantages in both computational cost and performance in the literature \cite{Hatomura1}.

In examining quantum circuit complexities for simulating honeycomb lattice configurations within the FH model, a distinct variation emerges between adiabatic and non-adiabatic protocols with CD terms. In conventional adiabatic evolution, the complexity depends linearly on the lattice site count, $N_{\text{site}}$. On the contrary, non-adiabatic evolution assisted by CD terms incurs an additional complexity order of $\mathcal{O}(N_{\text{site}})$, effectively doubling the gate count relative to adiabatic evolution. This escalation in complexity, while facilitating expedited convergence to the system's ground state, poses increased demands on quantum computational resources. Therefore, the trade-off between simulation speed and quantum gate numbers requires careful consideration, particularly in the context of gate fidelity and operational coherence times.

\begin{figure}[t]
\centering
\includegraphics[scale=0.11]{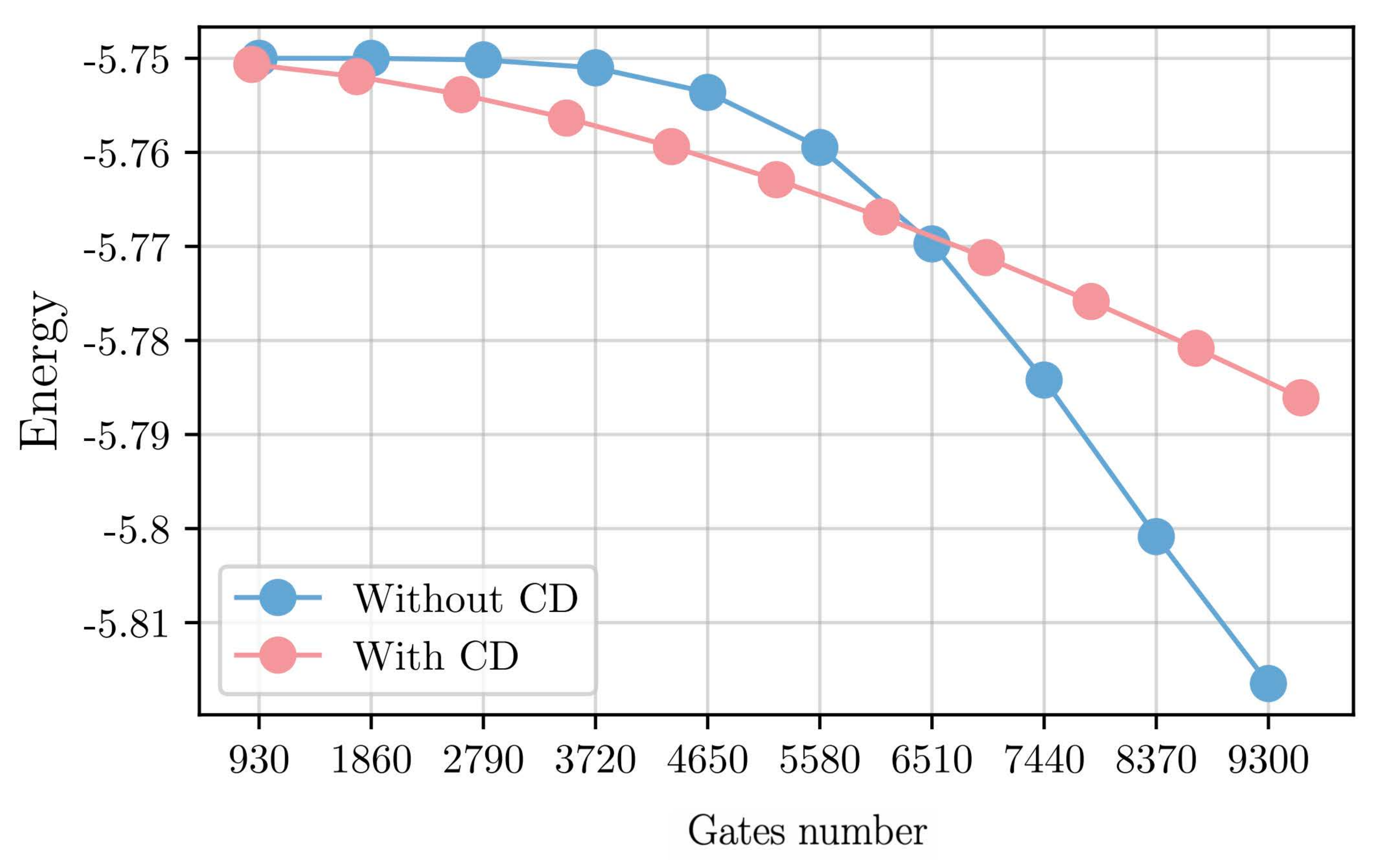}
\caption{Energy as a function of gate numbers. For a fair comparison, the numbers of basic quantum gates for adiabatic and CD-assisted non-adiabatic evolution are set to the same at each data point. Each Trotter step costs 310 basic gates without CD and 930 gates with CD.}
\label{fig4}
\end{figure}

Moreover, we have compared adiabatic and CD-assisted non-adiabatic evolution at the same gate number, as depicted in Fig.~\ref{fig4}. In our analysis, we find that within the range of 7 Trotter steps, comprising approximately 6500 quantum gates, counter-diabatic (CD) driving outperforms adiabatic evolution. However, as the number of quantum gates increases, the Trotter error surpasses the adiabatic error, leading to the inefficiency of CD. This reduced efficiency of the CD method at equivalent depths is due to its inherently higher complexity, which negatively affects its performance compared to the adiabatic approach when both are subjected to the same number of gate constraints. This suggests that there is potential for further investigation into the optimal compilation process to reduce the number of CNOT gates and consequently the complexity of the circuits~\cite{ji}.

Rather than implementing the entire Hamiltonian specified in Eq.~\eqref{eq:cd} within quantum circuits, we finally check the evolution of Hamiltonian with only CD term. As illustrated in Fig.~\ref{fig3} (a)-(b), considering only CD terms initially accelerates the system. However, over time, it fails to reach the ground state compared to cases involving adiabatic or non-adiabatic protocols assisted with CD terms. This discrepancy arises from the fact that approximate CD terms introduce large adiabatic error as well as non-negligible Trotter error \cite{Hatomura1}, while exact CD driving reproduces the dynamics of $\hat{H}_{\text{tot}}$ \eqref{eq:cd}, up to a phase factor \cite{Berry}, without any adiabatic error.  Therefore, this observation provides further insight into the trade-off between circuit depth (complexity) and errors (accuracy).

\begin{figure*}[t]
\centering
\includegraphics[scale=0.55]{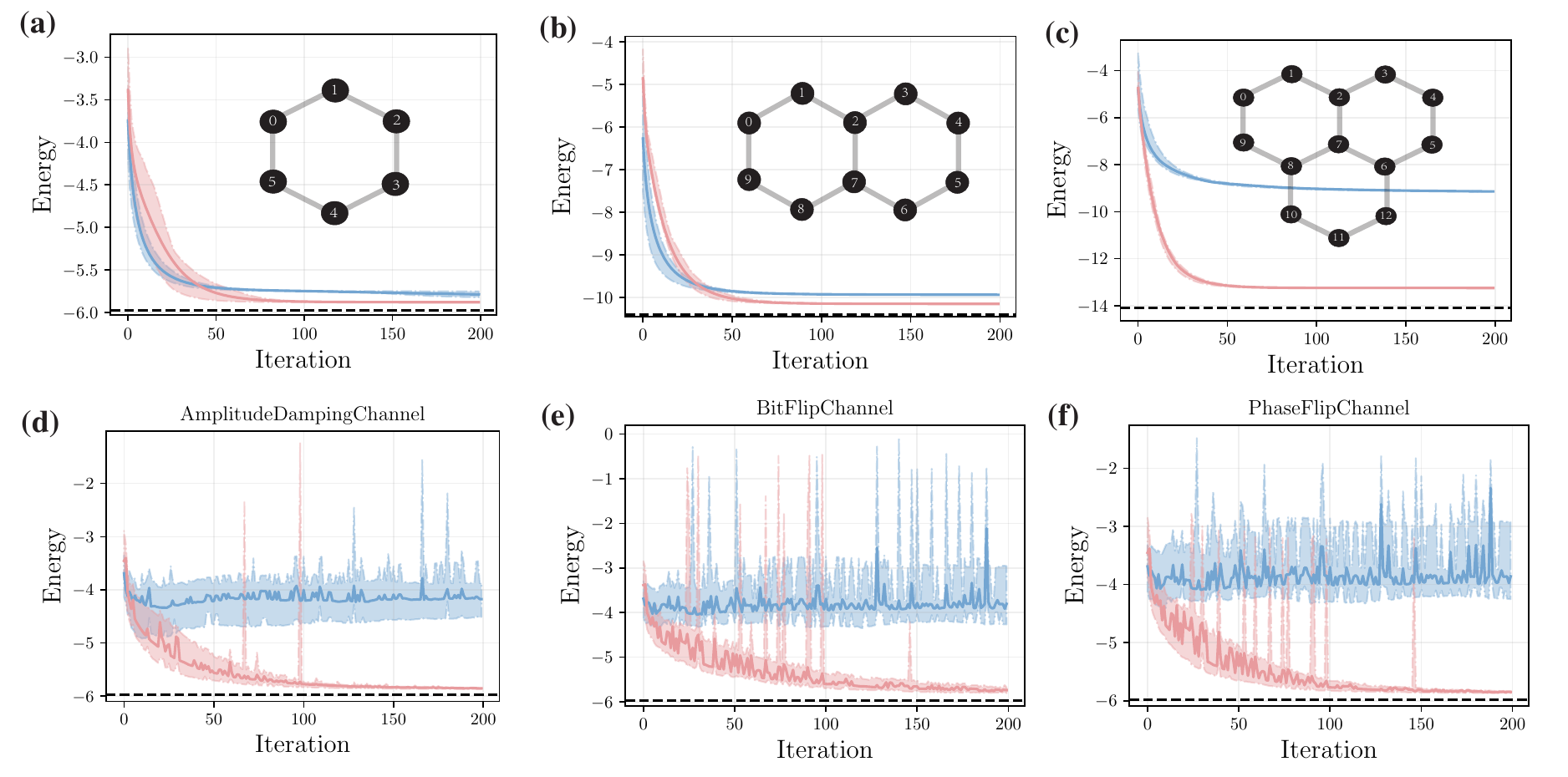}
\caption{Energy as a function of iteration. The blue solid and red solid lines show the results from HV ansatz and CD-inspired ansatz, respectively. The black dashed line shows the exact ground energy for comparison. Shaded regions present the upper and lower quartiles of ten instances. (a)-(c): System's energy corresponding to $1\times 1$, $1\times 2$ and $2 \times 1$ honeycomb lattice structure. The ground states vary in different cases. (d)-(f): The influence of amplitude damping noise, bit flip noise, and phase flip noise on the energy, exemplified by a $1\times 1$ honeycomb lattice structure. The other parameters used are $\tau=1$ and  $U=1.5$.}
\label{fig5}
\end{figure*}

\section{\label{sec:level4} Variational Quantum Algorithm}

Unlike AQAs, which fix parameters, VQAs optimize parameters using classical optimizers, for instance, the Adagrad optimizer used here. This distinction allows for the implementation of more efficient and shallower quantum circuits, facilitating the determination of ground state energy within the reduced evolution time. 
Therefore, in this section, we turn to the study of VQAs for our proposal with CD interaction.

In general, the quantum state $|\psi(\boldsymbol{\theta})\rangle$ is generated by applying the unitary operator to initial state, e.g., $|\psi(\boldsymbol{\theta})\rangle = U(\boldsymbol{\theta})|\psi_0\rangle$, where $U(\boldsymbol{\theta})$ can be implemented by quantum circuits and $\boldsymbol{\theta}=(\theta_0,~\theta_1,~\cdots,~\theta_m)$ are the parameters that need to be optimized.
The cost function can be defined as 
\begin{equation}
C(\boldsymbol{\theta}) = \langle \psi(\boldsymbol{\theta})|H_{\text{p}}|\psi(\boldsymbol{\theta})\rangle,
\end{equation}
where $H_{\text{p}}$ is the problem Hamiltonian. The VQA algorithm aims to obtain the
optimal parameters $\boldsymbol{\theta}^*$, which will minimize the cost function and ultimately result in achieving the lowest energy possible.
In what follows we will introduce two types of variational ansatz: Hamiltonian variational (HV) ansatz and CD-inspired ansatz, as suggested by DAQAs. \cite{Chandarana2022PRR, Pranav2023PRA, Xu2024Arxiv}

\subsection{Hamiltonian variational ansatz}

The HV ansatz is inspired by the adiabatic theorem, which explains that a sufficient slow evolution will maintain the ground state of the system~\cite{Wecker2015PRA}. The quantum circuits utilized here closely follow the method proposed in Ref. \cite{ObiolPRA2022}, serving as essential components in the implementation of adiabatic evolution. However, a notable challenge arises in the current landscape of quantum computing, particularly in implementing the vertical terms of the hopping Hamiltonian across non-adjacent qubit locations.

To tackle the complexity associated with non-adjacent qubits in the hopping terms, a key element — the Fermi-SWAP gate is introduced to transform them into adjacent qubits. This gate proves to be a crucial tool for efficiently swapping qubits, thereby transforming non-adjacent qubit interactions into adjacent ones. More importantly, the incorporation of a phase factor with a value of $-1$ ensures the preservation of Fermi properties during the interchange of two-qubit indices. Alongside this, the matrix representation of the Fermi-SWAP gate is given by:
\begin{align}
{\rm FSWAP}
&=
\left(
\begin{array}{cccc}
1 & 0  & 0  & 0\\
0  & 0  & 1 & 0 \\
0 & 1  & 0  & 0\\
0 & 0  & 0  & -1
\end{array}
\right),
\end{align}
which facilitates the exchange.
 
For the hopping Hamiltonian ($\hat{H}_{\text{h}}$), when the qubits involved in the hopping term are adjacent, the evolution operator and its matrix representation can be depicted as follows:
\begin{align}
U_{\text{h}}(\beta) =  e^{-i \beta \hat{\Sigma}_{\text{h}}}
= \begin{pmatrix}
1 & 0  & 0  & 0\\
0  & \cos\beta  & -i\sin \beta & 0   \\
0 & -i\sin \beta  & \cos \beta  & 0\\
0 & 0  & 0  & 	1
\end{pmatrix},
\label{eq: Uh}
\end{align}
where $\hat{\Sigma}_{\text{h}}= \frac{1}{2}\left(X_0X_1 + Y_0Y_1 \right)$. Given that the qubit indices for each Coulomb term ($\hat{H}_\text{c}$) are adjacent, representing two spins at the same site, SWAP operations are unnecessary. Therefore, the matrix representation 
of its evolution operator $U_{\text{c}}(\gamma)$ can be written as
\begin{align}
U_{\text{c}}(\gamma) =  e^{-i\gamma \hat{\Sigma}_\text{c}}
= \begin{pmatrix}
1 & 0  & 0  & 0\\
0  & 1  & 0 & 0   \\
0 & 0  & 1  & 0\\
0 & 0  & 0  & e^{-i\gamma} 
\end{pmatrix},
\label{eq: Uc}
\end{align}
where $\hat{\Sigma}_{\text{c}} = \frac{1}{4}\left(I-Z_0-Z_1+Z_0Z_1 \right)$.  Both parameters $\beta$ and $\gamma$ are determined by the classical optimizer, as they are essential components of VQAs.

The circuits corresponding to HV ansatz are depicted in Fig.~\ref{fig2}(c), where the Coulomb and hopping terms are present in the circuit. The Fermi-SWAP operation acts on the circuit and modifies the order of qubits. In pursuit of enhanced flexibility for optimization, we adopt a strategy wherein every Pauli operator is encoded with a free parameter.

\subsection{CD-inspired ansatz}


The structure of the CD-inspired ansatz is expressed as follows:
\begin{equation}
U_{\text{CD}}(\bm{\theta}) = e^{-i\bm{\theta}\hat{\Sigma}{\text{c}}},
\end{equation}
where $\hat{\Sigma}{\text{c}}$ are obtained from the NC method, corresponding to the gauge potential $\hat{A}_\lambda$. \XC{In general, combining Hamiltonian variational ansatz with CD-inspired ansatz in VQAs is possible, but it often leads to a significant increase in circuit depth and complexity. Since CD driving can mimic the full dynamics of the Hamiltonian, up to the dynamic phase \cite{Berry}, we can exclusively employ CD-inspired ansatz for our purposes.} This offers the significant benefit of eliminating many terms from the original Hamiltonian, allowing for the implementation of lower-depth circuits compatible with near-term quantum devices. This approach utilizes operators selected from the CD pools, with each term incorporating its own set of free parameters. These parameters are subject to optimization via a classical optimizer, facilitating an efficient exploration of the parameter space to achieve optimal simulation results. 

To achieve this, the first-order CD operator pool needs to be computed, including two-body interactions, four-body interactions, and many-body interactions. In the VQA algorithms, we solely include the two-body interactions as fundamental and feasible elements. These chosen terms, denoted as $X_0Y_2, Y_2X_0, \ldots, X_9Y_{11}, Y_{11}X_9$, correspond to specific Pauli terms, representing the crucial interactions within the system. A schematic representation of the CD-inspired ansatz utilized for the FH model is depicted in Fig.~\ref{fig2}(d), illustrating the structure with two-body interactions.

\subsection{Classical optimizer}

In VQA, the optimization of parameters is crucial and is accomplished by a classical computer aiming to minimize a cost function, thereby identifying the optimal set of parameters. Indeed, various optimizers, such as Adam, Adagrad, and SGD can also be utilized for various tasks, each offering distinct advantages and disadvantages \cite{Xu2024Arxiv}. Therefore, the choice of optimizers depends on factors such as the nature of the optimization problem, the structure of the neural network, and computational resources available.

The Adagrad optimizer, a widely recognized optimization technique in the realms of machine learning and deep learning, offers an efficient approach to parameter optimization by dynamically adjusting the learning rate for each parameter, as detailed in ~\cite{duchi2011adaptive}. The core principle of Adagrad lies in its mechanism to monitor and accumulate the squared gradients for each parameter over time. This accumulation process is particularly advantageous in handling sparse data and scenarios where the gradient magnitude varies significantly across parameters. By adapting learning rates on an individual parameter basis, Adagrad emerges as a robust and versatile optimization algorithm, especially in contexts where parameter sensitivities differ markedly. 
Mathematically, the Adagrad optimizer computes the updates for the parameters using the following formulas:
\begin{equation}
    \bm{\theta}_{t+1} = \bm{\theta}_{t} - \frac{\eta}{\sqrt{G_{t} + \epsilon}} \cdot \nabla J(\bm{\theta}_{t}),
\end{equation}
where $\eta$ is the global learning rate, $G_{t}$ is the accumulated squared gradient for the parameter $\bm{\theta}$ up to iteration $t$, and $\epsilon$ is a small constant added for numerical stability.

\subsection{Simulation}

To evaluate the performance of VQA in determining the ground energy of the FH model structured on a honeycomb lattice, we conduct a comparative analysis of the HV ansatz and CD-inspired ansatz. For a fair comparison, we utilize quantum circuits consisting of a single layer for each type of ansatz. Employing the Adagrad optimizer, known for its effectiveness in adapting learning rates based on the parameter's gradient history, we set the learning rate ($\eta$) at $0.05$. Due to the significant influence of initial parameters on the outcomes, we initialize these parameters using a uniform distribution within the interval $(0, 1)$. This random initialization as the starting point ensures an unbiased assessment of each ansatz's ability to efficiently approximate the ground state energy of the model.

In simulations involving $1\times 1$, $1\times 2$, and $2\times1$ honeycomb lattices requiring 12, 20, and 26 qubits respectively, the performances of HV ansatz and CD-inspired ansatz are compared. Results, as depicted in Fig.~\ref{fig5}(a)-(c), indicate that CD-inspired ansatz yields a closer approximation to the ground state energy than the HV ansatz. Furthermore, the CD-inspired ansatz achieves convergence significantly faster than the HV ansatz. When subjected to various noise sources, such as amplitude damping, bit flip, and phase flip noise, each with a probability of $0.01$, the HV ansatz encounters challenges in achieving convergence, as depicted in Fig.~\ref{fig5}(d)-(f). On the contrary, the CD-inspired ansatz exhibits robustness, maintaining convergence even in the presence of noise. This underscores its potential effectiveness in noisy quantum computing environments.
 
\begin{figure*}[t]
\centering
\includegraphics[scale=0.5]{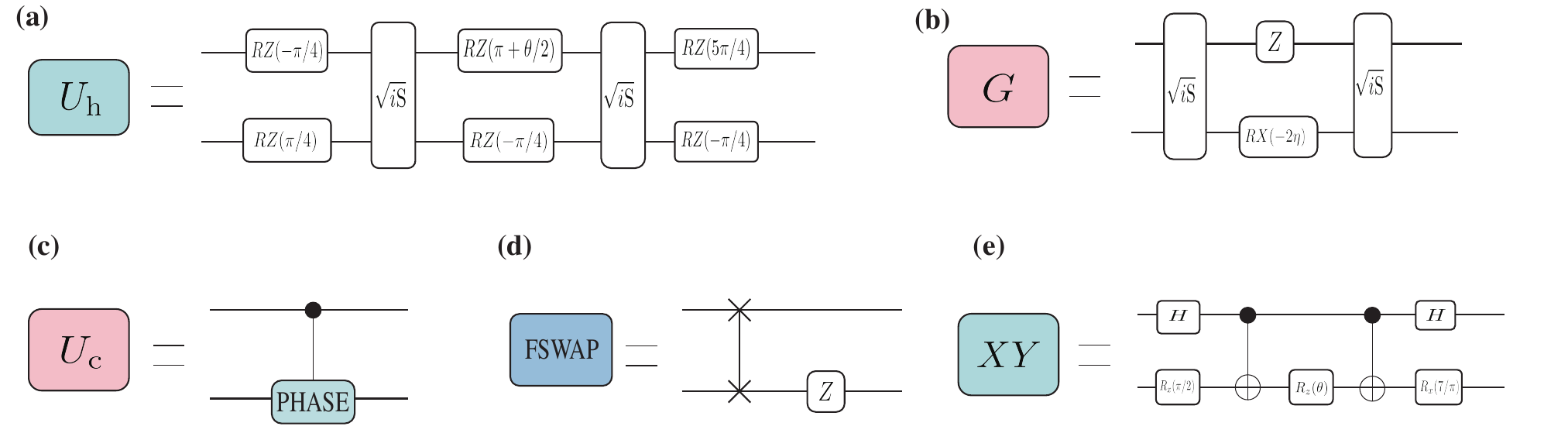}
\caption{Evolution operators used in the quantum circuits. (a): the hopping operator. (b): the Givens rotation operator. (c): the Coulomb operator. (d): Fermi-SWAP Operator. (e): XY operator. }
\label{fig6}
\end{figure*}

Next, we compare the quantum circuit complexities across different honeycomb lattice configurations. The number of two-body interactions in the CD-inspired ansatz directly correlates with the number of grid points, denoted as $N_{\text{CD}} = 2N_{\text{site}} - 4$. Given that each two-body operator XY can be decomposed into $7$ basic quantum gates shown in Fig.~\ref{fig6}(e), the total number of quantum gates for the CD-inspired ansatz is approximately $\mathcal{O}(N_{\text{CD}})$. On the other hand, for the HV ansatz, the quantum circuit's architecture consists of three main components: the hopping evolution operator, the Coulomb evolution operator, and the fermionic SWAP unit. This structure remains consistent across various honeycomb lattice sizes, with the total quantum gate count adhering to the formula $N_{\text{HV}} = N_{\text{Coulomb}} + N_{\text{hopping}} + N_{\text{swap}}$. The gate count for each component is inherently linked to the number of lattice sites. Specifically, $N_{\text{Coulomb}}$ equals the number of sites $N_{\text{site}}$, $N_{\text{hopping}}$ is twice the number of sites $2N_{\text{site}}$, and $N_{\text{swap}}$ is a function of the lattice configuration, calculated as $2\sum_{i\in x}^{y} \left( N_{\text{site}}^{i}(N_{\text{site}}^{i}/2 + (N_{\text{site}}^{i}-2)/2)\right)$. Here, the SWAP operation counts, $N_{\text{site}}^{i}/2$ and $(N_{\text{site}}^{i}-2)/2$, account for two distinct types of SWAP gates required within the lattice.

As the size of the honeycomb lattice increases, the scaling of circuit complexity between the CD-inspired ansatz and the HV ansatz diverges significantly. In the case of CD-inspired ansatz, the complexity scales linearly with the number of sites, denoted as \(\mathcal{O}(N_{\text{site}})\), reflecting a direct correlation between lattice size and resource requirements. Conversely, for the HV ansatz, the complexity exhibits exponential growth, represented as \(\mathcal{O}(N_{\text{site}}^2)\), due to the quadratic increase in SWAP operations as the lattice expands. This different scaling behavior implies CD-inspired ansatz is more efficient for larger lattices, whereas the HV approach has the dilemma of requesting a large demand of quantum gates.

\XC{
Finally, let us compare the results shown in Figs. \ref{fig3} and \ref{fig5}, highlighting the advantages and disadvantages of VQAs and AQAs, both assisted by CD driving. Due to the random initialization, the search space in VQA is enlarged, potentially leading to suboptimal results compared to adiabatic evolution. Consequently, the ground energy calculated from VQAs becomes inferior. On the other hand, AQAs demand significant computational resources, especially when simulating large-scale quantum systems. For example, through our estimation, lattice configurations like $1\times 2$ and $2\times 1$ require approximately 50,000 quantum gates, while a $1\times 3$ lattice needs about 60,000 quantum gates (up to 26 qubits), even without CD interactions involved. These challenges render the applicability of AQAs impractical on near-term quantum devices.}

\section*{\label{sec:level5} CONCLUSION and Outlook}

In summary, our study detailed in the paper has utilized CD interaction to compute the ground energy of the honeycomb lattice within the FH model, using both AQAs and VQAs. To facilitate the simulation of one-dimensional ($1 \times 1$ and $1 \times 2$) and two-dimensional ($2 \times 1$) honeycomb lattices up to 26 qubits with manageable circuit depth, we have introduced the CD-inspired ansatz, drawn from the concept of shortcuts to adiabaticity. This choice, different from conventional HV ansatz, has exhibited superior performance, particularly under various noise conditions. These findings not only contribute to the accurate calculation of quantum material energy within specific regions but also suggest potential applications in exploring properties of materials such as artificial graphene \cite{ObiolPRA2022}, high-temperature superconductors \cite{dong} and quantum dots \cite{wang}, possibly with the extended FH model.

In addition, we have demonstrated the efficacy of our approaches compared to traditional AQAs and VQAs in accurately determining ground state energy. Comprehensive analysis and error evaluation suggest that the performance of both algorithms is improved by CD driving, allowing for a reduction in circuit depth and consequently a decrease in the number of gates required. More specifically, the introduction of CD-inspired ansatz within the VQA framework improves effectiveness in accurately approximating ground state energies by reducing circuit depth and complexity, especially under various noise conditions. Although the accuracy in the case of a $1\times 1$ honeycomb lattice structure is worse than that of the AQA, the CD-inspired ansatz offers a promising avenue for mitigating challenges associated with large-scale qubits. The latter allows us to obtain the ground state of a $2 \times 1$ honeycomb lattice structure with certain accuracy and shallower quantum circuits, using systems with up to 26 qubits.

Indeed, AQAs and VQAs have their own pros and cons. Our results indicate that VQAs even with CD-inspired ansatz suffer the disadvantage, such as sensitivity to initialization and barren plateaus \cite{Larocca}, commonly encountered in optimization landscapes. 
Unlike AQAs, VQAs lack rigorous theoretical guarantees of convergence to the ground state, making their performance highly dependent on the choice of ansatz and optimization method. Moving forward, these suggest that Krylov space \cite{krylov} could be useful for selecting the ansatz, particularly for
large-scale quantum systems. Despite these challenges, we remain hopeful that with advancements in quantum hardware capabilities, our methods could play a crucial role in the comprehensive study of a broader range of materials, offering enhanced insights into their quantum mechanical properties and potentially contributing to the development of new materials with desirable electronic characteristics.

\begin{acknowledgments}
This work is supported by the Basque Government through Grant No. IT1470-22, the project grant PID2021-126273NB-I00 funded by MCIN/AEI/10.13039/501100011033, by ``ERDFA way of making Europe", ``ERDF Invest in your Future",  EU FET Open Grant EPIQUS (899368), 
HORIZON-CL4-2022-QUANTUM-01-SGA project 101113946 OpenSuperQPlus100 of the EU Flagship on Quantum Technologies,
the Spanish Ministry of Economic Affairs and Digital Transformation through the QUANTUM ENIA project call-Quantum Spain project, NSFC (12075145 and 12211540002), the Innovation Program for Quantum Science and Technology (2021ZD0302302), and
the China Scholarship Council (CSC) under Grant Nos: 202206890003, 202306890004. Y.B. acknowledges support from the Spanish Government via the project PID2021-126694NA-C22 (MCIU/AEI/FEDER, EU). G.P. was supported
by Spain’s MINECO through Grant No. PID2020-
117787GB-I00 and by Spanish National Research Council
(CSIC) Research Platform PTI-001. X.C. acknowledges ayudas para contratos Ram\'on y Cajal–2015-2020 (RYC-2017-22482).
\end{acknowledgments}

\appendix
\section{Calculation of CD terms}
\label{app: cd pool}

Within the CD operator pool for the FH model, NCs generate various terms, including two-body, four-body, and many-body interactions. While two-body interactions are less resource-intensive, incorporating many-body interactions corresponding to high-order NCs is crucial for enhancing simulation accuracy. This appendix delves into the specifics of NCs within the FH model, emphasizing their role in accurately capturing the system's dynamics. Despite the increased quantum resources required for many-body interactions, they are essential for a comprehensive simulation, providing a deeper understanding of the model's complex behaviors and interactions.

Now, we only consider two sites with qubit index $(0,2)$ and $(1,3)$. The Hamiltonian is
\begin{equation}
    H = a^\dagger_0a_0a^\dagger_1 a_1+a^\dagger_2a_2a^\dagger_3 a_3
    + a^\dagger_0a_2 + a^\dagger_2a_0 + a^\dagger_1a_3 + a^\dagger_3a_1.
\label{B1}
\end{equation}
With JW transformation, the creation and annihilation operator will be transformed into a qubit operator, such that
\begin{eqnarray}
H_{\text{c}} &=& 2I-Z_0-Z_1-Z_2-Z_3+Z_0Z_1+Z_2Z_3, \\
H_{\text{h}} &=& 0.5(Y_0Z_1Y_2 + X_0Z_1X_2 + Y_1Z_2Y_3 + X_1Z_2X_3).
\end{eqnarray}
Since the initial Hamiltonian is chosen as the hopping Hamiltonian, the NC, is given by the hopping Hamiltonian and Coulomb Hamiltonian, that is,
\begin{align}
H_{\text{CD}} &= [H_{\text{c}}, H_{\text{h}}]\\
&= X_0Y_2 + Y_0X_2 + X_0Z_1X_2Z_3 + Y_0Z_1X_2Z_3 \nonumber \\
&+ X_1Y_3 + Y_3X_1 + Z_0X_1Z_2Y_3 + Z_0Y_1Z_2X_3, \nonumber
\end{align}
When incorporating vertical hopping terms into the hopping Hamiltonian, the resulting CD Hamiltonian ($H_{\text{CD}}$) inevitably encompasses higher-order interaction terms. This complexity arises from the necessity to counteract the non-adiabatic transitions induced by these additional hopping interactions.

The JW transformation, when applied to this expanded Hamiltonian, retains the intrinsic relationships between fermionic operators. This transformation effectively maps the fermionic creation and annihilation operators to Pauli matrices, preserving the algebraic structure of the fermionic system within a spin-based framework. Consequently, when computing the NCs from the original Hamiltonian ~(\ref{B1}), the JW transformation ensures that the outcome remains consistent, reflecting the preserved fermionic operator relationships.

\section{Digitized quantum circuits}
\label{app: digital}

In this appendix, we will introduce how to digitalize adiabatic quantum computing. For a system whose Hamiltonian is $H(t)$ undergoes an evolution, the continuous time evolution operator can be formulated by 
\begin{equation}
{U}(0, T) = \mathcal{T}{\rm exp} \left[-i\int_0^T dt \hat{H}(t)\right],
\end{equation}
where the $\mathcal{T}$ is the time-ordering operator. The discretization of the evolution operator can be designed using the Trotter-Suzuki approximation 
\begin{equation}
U(0,T) \approx \prod_{j=1}^p\prod_{m=1}^{M} {\rm exp}\{-iH_{\text{m}}(j\Delta t)\Delta t \},
\end{equation}
where $H(t)$ can be decomposed into $M$ local terms and $p$ refers to the number of circuit layers. If $p$ is large enough, the evolution $U(0, T)$ can generate an exact target state. Discretizing the adiabatic evolution, as implemented in superconducting circuits~\cite{Barends_Nature2016}, also enables counter-adiabatic evolution. 

Fig.~\ref{fig6}(a)-(c) delineates the circuit decomposition for various evolution operators discussed within this work. Specifically, for the hopping evolution operator \(U_{\text{h}}(\beta)\), the quantum circuits incorporate rotation \(Z\) gates and \(\sqrt{\rm iSWAP}\) gates, illustrating the implementation strategy for hopping interactions. In contrast, the Coulomb evolution operator \(U_{\text{c}}(\gamma)\) extends the circuit complexity by including both rotation \(X\) and \(Z\) gates alongside \(\sqrt{\rm iSWAP}\) gates, reflecting the additional requirements for simulating Coulomb interactions within the quantum framework.

The construction of the Fermi-SWAP operator, as shown in part (d) of Fig.~\ref{fig6}, employs rotation \(Z\) gates and \(\sqrt{\rm iSWAP}\) gates, underscoring the methodology for effectuating fermion swapping within the lattice simulation. Lastly, Fig.~\ref{fig6}(e) also exemplifies the decomposition of two-body interaction terms specific to the CD-inspired ansatz, such as the operator \(e^{-iX_{0}X_{2}t/\hbar}\). This decomposition utilizes basis quantum gates, including the Hadamard gate, Rotation gate, and CNOT gate, to effectively simulate the desired interaction within the quantum circuit framework.

This detailed decomposition showcases the versatility and complexity of quantum circuit design necessary to accurately simulate the dynamic interactions characteristic of the FH model on a honeycomb lattice, facilitating a deeper understanding of quantum material properties.

\end{document}